\pdfoutput=1
\PassOptionsToPackage{table,xcdraw, svgnames}{xcolor}
\documentclass[sigconf]{acmart}
\setcopyright{none}
\renewcommand\footnotetextcopyrightpermission[1]{} 
\usepackage{draftwatermark}

\usepackage{subfigure}
\usepackage{verbatim}
\usepackage{RPMTools}
\usepackage{tabu}                      
\usepackage{booktabs}                  
\newcommand\cut[1]{}
\AtBeginDocument{%
  \providecommand\BibTeX{{%
    \normalfont B\kern-0.5em{\scshape i\kern-0.25em b}\kern-0.8em\TeX}}}

\begin{document}

\title[VALID: A perceptually validated Virtual Avatar Library for Inclusion and Diversity]{VALID: A perceptually validated Virtual Avatar Library for Inclusion and Diversity}


\author{Tiffany D. Do}
\email{tiffanydo@knights.ucf.edu}
\affiliation{%
  \institution{University of Central Florida}
  \city{Orlando}
  \state{Florida}
  \country{USA}
}

\author{Steve Zelenty}
\email{steve.zelenty@knights.ucf.edu}
\affiliation{%
  \institution{University of Central Florida}
  \city{Orlando}
  \state{Florida}
  \country{USA}
}

\author{Mar Gonzalez-Franco}
\email{margon@google.com}
\affiliation{%
  \institution{Google Labs}
  \city{Seattle}
  \state{Washington}
  \country{USA}
}

\author{Ryan P. McMahan}
\email{rpm@ucf.edu}
\affiliation{%
  \institution{University of Central Florida}
  \city{Orlando}
  \state{Florida}
  \country{USA}
}

\SetWatermarkText{PREPRINT}
\SetWatermarkLightness{0.9}
\SetWatermarkScale{1}

\renewcommand{\shortauthors}{Do et al.}

\begin{abstract}
  As consumer adoption of immersive technologies grows, virtual avatars will play a prominent role in the future of social computing. However, as people begin to interact more frequently through virtual avatars, it is important to ensure that the research community has validated tools to evaluate the effects and consequences of such technologies. We present the first iteration of a new, freely available 3D avatar library called the \textit{Virtual Avatar Library for Inclusion and Diversity} (\textit{VALID}), which includes 210 fully rigged avatars with a focus on advancing racial diversity and inclusion. We present a detailed process for creating, iterating, and validating avatars of diversity. Through a large online study ($n=132$) with participants from 33 countries, we provide statistically validated labels for each avatar's perceived race and gender. Through our validation study, we also advance knowledge pertaining to the perception of an avatar's race. In particular, we found that avatars of some races were more accurately identified by participants of the same race.
\end{abstract}

\maketitle
\pagestyle{plain}
\section{Introduction}
The study of human behavior in virtual worlds is becoming increasingly important as more and more people spend time in these digital spaces. In particular, understanding how behaviors in virtual environments compare to behaviors in real life can provide insight into the ways in which humans interact with technology and each other. Furthermore, it can also help inform the design of virtual environments, making them more realistic and socially acceptable.

One area of significant interest is virtual avatars, which are 3D representations of virtual humans often used in virtual worlds and simulations \cite{Pelachaud2009}. Virtual avatars have a prominent role in social computing and immersive environments \cite{YeeBailenson2007, Steinicke2017, Fox2015}. For example, previous studies have utilized virtual avatars to express emotions \cite{Butler2017}, recreate social psychology phenomena \cite{gonzalez2018participant, Do2022}, and influence feelings of presence or embodiment \cite{Peck2021, Nowak2003}. Moreover, virtual avatars are expected to play evermore important roles in the future. For instance, virtual avatars are at the center of every social virtual reality (VR) and augmented reality (AR) interaction, as they are key to representing remote participants \cite{Piumsomboon2018} and facilitating collaboration \cite{Sra2015}.

Open virtual avatar libraries facilitate research by providing freely available resources to the community. For instance, Microsoft's Rocketbox avatar library \cite{Gonzalez2020} has been extensively used in multiple studies (e.g., \cite{Cheng2022, Jeong2020, Huang2022}), with over 400 stars and 100 forks on GitHub since its release in 2020. However, the representation of races in currently available avatar libraries are limited and have not been validated through user perception studies. As examples, over 40\% of the Rocketbox avatars are white males, and MakeHuman\footnote{http://www.makehumancommunity.org/}, another popular resource for 3D avatars, only includes options for African, Asian, and Caucasian avatars. Due to such limitations, many researchers are unable to properly investigate the effects of diversity for virtual avatars. Furthermore, existing libraries not only have limited diversity but are also biased in their lack of inclusivity with regard to professions for minority avatars. For instance, Rocketbox lacks Asian avatars dressed for medical scenarios.

The limited representation of currently available resources may have an inherently detrimental impact by introducing biases into research investigations from the outset, such as restricting the development of scenarios with Asian medical professionals. Furthermore, as immersive technologies grow in popularity worldwide, Taylor et al. \cite{Jones2020} have called for improved racial/ethnic representations in VR to mitigate potential negative effects of racial bias. This can be especially important since virtual avatar race\footnote{We use the term race to refer to a ``visually distinct social group with a common ethnicity" \cite{Rhodes2010}.} can have a strong influence on human behavior. For example, research has found that embodying a virtual avatar of a different race can affect implicit racial bias, as demonstrated by several prior studies \cite{Peck2013, groom2009, Maister2015, Salmanowitz2018}. As such, it is critical to provide more inclusive avatar resources.

The goal of this work was dual purpose. First, we sought to develop an open library of validated virtual avatars, which we refer to as the \textit{Virtual Avatar Library for Inclusion and Diversity} (\textit{VALID}). This library has been designed to inclusively represent a range of races across various professions. Secondly, we wanted to better understand how humans perceive avatars, particular with regard to race. In this first iteration of VALID, we followed the recommendations of the 2015 National Content Test Race and Ethnicity Analysis Report from the U.S. Census Bureau \cite{Matthews2017} to ensure that a wide range of people was represented in our library. For example, this report found that members of Middle Eastern and North African (MENA) communities use the MENA category when it is available, but have trouble identifying their race when it is not available. Similarly, it also found that Hispanics can better identify their ethnicity by combining the conventional race and Hispanic origin questions into one. Therefore, to overcome those traditional problems on racial and ethnic categorization, VALID includes seven races as recommended by the U.S. Census Bureau report \cite{Matthews2017} (which differs from the 2020 U.S. Census): American Indian or Native Alaskan (AIAN)\footnote{We use racial abbreviations as defined in the U.S. Census Bureau report.}, Asian, Black or African American (Black), Hispanic, Latino, or Spanish (Hispanic), Middle Eastern or North African (MENA), Native Hawaiian or Pacific Islander (NHPI), and White.

VALID includes 210 fully rigged virtual avatars designed to advance diversity and inclusion. We iteratively created 42 base avatars\linebreak(7 target races $\times$ 2 genders $\times$ 3 individuals) using a process that combined data-driven average facial features with extensive collaboration with representative stakeholders from each racial group. To address the longstanding issue of the lack of diversity in virtual designers and to empower diverse voices \cite{boberg_designing_2008}, we adopted a participatory design method. This approach involved actively involving individuals ($n=22$) from diverse backgrounds, particularly different racial and ethnic identities, in the design process. By including these individuals as active participants, we aimed to ensure that their perspectives, experiences, and needs were considered and incorporated into the design of the avatars. 

Once the avatars were created, we sought to evaluate their perception on a global scale. We then conducted a large online study ($n=132$) with participants from 33 countries, self-identifying as one of the seven represented races, to determine whether the race and gender of each avatar are recognizable, and therefore validated. We found that all Asian, Black, and White avatars were universally identified as their modeled race by all participants, while our AIAN, Hispanic, and MENA avatars were typically only identified by participants of the same race, indicating that participant race can bias perceptions of a virtual avatar's race. We have since modeled the 42 base avatars in five different outfits (casual, business, medical, military, and utility), yielding a total of 210 fully rigged avatars. 

To foster diversity and inclusivity in virtual avatar research, we are making all of the avatars in our library freely available to the community as open source models. In addition to the avatars, we are also providing statistically validated labels for the race and gender of all 42 base avatars. Our models are available in FBX format, are compatible with previous libraries like Rocketbox \cite{Gonzalez2020}, and can be easily integrated into most game engines such as Unity and Unreal. Additionally, the avatars come equipped with facial blend shapes to enable researchers and developers to easily create dynamic facial expressions and lip-sync animations. All avatars, labels, and metadata can be found at our GitHub repository: 
\url{https://github.com/xrtlab/Validated-Avatar-Library-for-Inclusion-and-Diversity---VALID}.

This paper makes three primary contributions: 

\begin{enumerate}
    \item We provide 210 openly available, fully rigged, and perceptually validated avatars for the research community, with a focus on advancing diversity and inclusion.
    \item Our diversity-represented user study sheds new light on the ways in which people's own racial identity can affect their perceptions of a virtual avatar's race. In our repository, we also include the agreement rates of all avatars, disaggregated by every participant race, which offers valuable insights into how individuals from different racial backgrounds perceive our avatars.
    \item We describe a comprehensive process for creating, iterating, and validating a library of diverse virtual avatars. Our approach involved close collaboration with stakeholders and a commitment to transparency and rigor. This could serve as a model for other researchers seeking to create more inclusive and representative virtual experiences.
\end{enumerate}

\section{Related Work}
In this section, we describe how virtual avatars are used within current research in order to highlight the need for diverse avatars. We conclude the section with a discussion on currently available resources used for virtual avatars and virtual agents. 

\subsection{Effect of Avatar Race}
Virtual avatars are widely used in research simulations such as training, education, and social psychology. The race of a virtual avatar is a crucial factor that can affect the outcomes of these studies. For example, research has shown that underrepresented students often prefer virtual instructors who share their ethnicity \cite{Baylor2003, Moreno2006}. Similarly, studies have suggested that designing a virtual teacher of the same race as inner-city youth can have a positive influence on them \cite{Baylor2009}, while a culturally relevant virtual instructor, such as an African-American instructor for African-American children, can improve academic achievement \cite{Gilbert2008}. 

The design of virtual avatars is especially important for minority or marginalized participants. Kim and Lim \cite{KimLim2013} reported that minority students who feel unsupported in traditional classrooms develop more positive attitudes towards avatar-based learning. In addition, children with autism spectrum disorder treat virtual avatars as real social partners \cite{Li2021, Ali2020}. Therefore, to better meet the needs of all individuals participating in such studies, it is important for researchers to have access to diverse avatars that participants can comfortably interact with. Diversity in virtual avatars is important not only for improving representation, but also for enhancing the effectiveness of simulations. Halan et al. \cite{halan2015} found that medical students who trained with virtual patients of a particular race demonstrated increased empathy towards real patients of that race. Similarly, Bickmore et al. \cite{Bickmore2021} showed that interacting with a minority virtual avatar reduced racial biases in job hiring simulations. 

These findings highlight the importance of diverse and inclusive virtual avatars in research simulations and emphasize the need for more comprehensive representation of different races. Access to a wide range of validated avatars through VALID will help to create more inclusive and representative simulations, and enable researchers to investigate the impact of avatar race or gender on participants' experiences. This will help improve the inclusivity of simulations and contribute towards addressing issues of bias.

\subsection{Implicit Racial Bias and Virtual Avatars}
Avatars are becoming increasingly important in immersive applications, particularly in the realm of VR, where they are becoming ubiquitous \cite{Dewez2021}. Research suggests that the degree of similarity between a user's real body and their virtual avatar can influence embodiment, presence, and cognition \cite{Jo2017, Jung2018, McIntosh2020, Ogawa2020}. Maister et al. \cite{Maister2015} proposed that non-matching self-avatars can affect a user's self-association and influence social cognition. Moreover, studies have demonstrated that embodying a darker-skinned avatar in front of a virtual mirror can reduce implicit racial biases \cite{Peck2013, groom2009, Maister2015, Salmanowitz2018}, which are unconscious biases that can lead to discriminatory behavior \cite{chapman2013physicians}. For instance, Salmanowitz et al. \cite{Salmanowitz2018} found that a VR participant's implicit racial bias affects their willingness to convict a darker-skinned suspect based on inconclusive evidence. Similarly, Peck et al. \cite{Peck2021b} found that each participant's implicit racial bias was related to their nuanced head and hand motions in a firearm simulation. These foundational studies provide compelling evidence that embodying an avatar of a different race can affect implicit biases and further emphasize the need for diverse avatar resources.

Our study examines how participants perceive the race of diverse virtual avatars. While some studies have explored how a virtual avatar's race affects user interactions (e.g., \cite{Baylor2003, Gamberini2015, zipp2019impact}), little research has been conducted on how individuals \textit{actively perceive} the race of virtual avatars. Setoh et al. \cite{Setoh2019} note that racial identification can predict implicit racial bias, making it crucial to understand how people perceive the race of virtual avatars to further investigate these effects.

\subsection{Own-Race Bias}
Own-race bias, also known as the "other-race effect," refers to the phenomenon in which individuals process the faces of their own race differently from those of other races \cite{Rhodes2009, Civile2022, meissner2001, Ge2009}. Studies have suggested that this bias can influence the way individuals categorize race. For example, MacLin and Malpass \cite{MacLin2001} found that Hispanic participants were more likely to categorize Hispanic faces as fitting their racial category than Black faces, and Blascovich et al. \cite{Blascovich1997} observed that participants who strongly identify with their in-group are more accurate in identifying in-group members.

Although own-race bias has not yet been studied in the context of 3D virtual avatars, Saneyoshi et al. \cite{Saneyoshi2022} recently discovered that it extends to the uncanny valley effect \cite{mori2012} for 2D computer-generated faces. Specifically, they found that Asian and European participants rated distorted faces of their own race as more unpleasant than those of other races. Building on this research, we extended the study of own-race bias to 3D virtual avatars and focused on race categorization rather than perceived pleasantness. Our study included avatars and participants from seven different races, providing insights into how a diverse user population may interact within equally diverse virtual worlds.

\subsection{Virtual Avatar Resources}

There are numerous resources for creating virtual avatars. Artists can use 3D modeling tools, such as Autodesk 3ds Max\footnote{https://www.autodesk.com/products/3ds-max/overview}, Autodesk Maya\footnote{https://www.autodesk.com/products/maya/overview}, Blender\footnote{https://www.blender.org}, or ZBrush\footnote{https://www.maxon.net/en/zbrush} to manually model, texture, and rig virtual avatars. However, such work requires expertise in 3D modeling and character design, and is often a tedious process \cite{Gonzalez2020}. On the other hand, parametric models, including freely available tools like MakeHuman\footnote{http://www.makehumancommunity.org/} and Autodesk Character Generator\footnote{https://charactergenerator.autodesk.com/}, as well as commercially available ones such as Daz3D\footnote{https://www.daz3d.com}, Poser\footnote{https://www.posersoftware.com}, and Reallusion Character Creator\footnote{https://www.reallusion.com/character-creator/}, enable users to generate virtual avatars from predefined parameters, thereby significantly expediting the avatar generation process. Nonetheless, using these tools still requires learning a new program and time to customize each model, despite the absence of the artistic expertise needed for manual tools.

Another alternative to traditional modeling is to use scanning technologies, which can capture 3D models of real people. For instance, Shapiro et al. \cite{Shapiro2014} and Waltemate et al. \cite{Waltemate2018} used 3D cameras and photogrammetry, respectively, to capture 3D models of their users. Singular Inversions FaceGen Modeller\footnote{https://facegen.com/modeller.htm} has also been employed to generate 3D faces from user photos and then apply them to a general 3D avatar body \cite{blom2014achieving, gonzalez2016neurological}. However, scanning approaches require the ability to physically scan the user, limiting their use for certain applications, particularly remote ones.

Most closely related to our goal of providing a free and open library of ready-to-use avatars is the Microsoft Rocketbox library \cite{Gonzalez2020} and its accompanying HeadBox \cite{Volonte2022} and MoveBox \cite{MoveBox2020} toolkits. Rocketbox provides a free set of 111 fully rigged adult avatars of various races and outfits. However, it falls short in terms of representation by not including any avatars of AIAN or NHPI descent. Additionally, the library offers only a limited number of Asian, Hispanic, and MENA avatars, excluding minority representations for some professions (e.g., Rocketbox does not include any Asian medical avatars). Furthermore, none of the available avatar libraries have been validated by user perception studies to ensure their efficacy and inclusivity. Therefore, our VALID project aims to fill this gap by providing a free and validated library of diverse avatars.

\section{Avatar Creation Procedure}
\begin{figure*}[h!]
    \centering
    \includegraphics[width=7in]{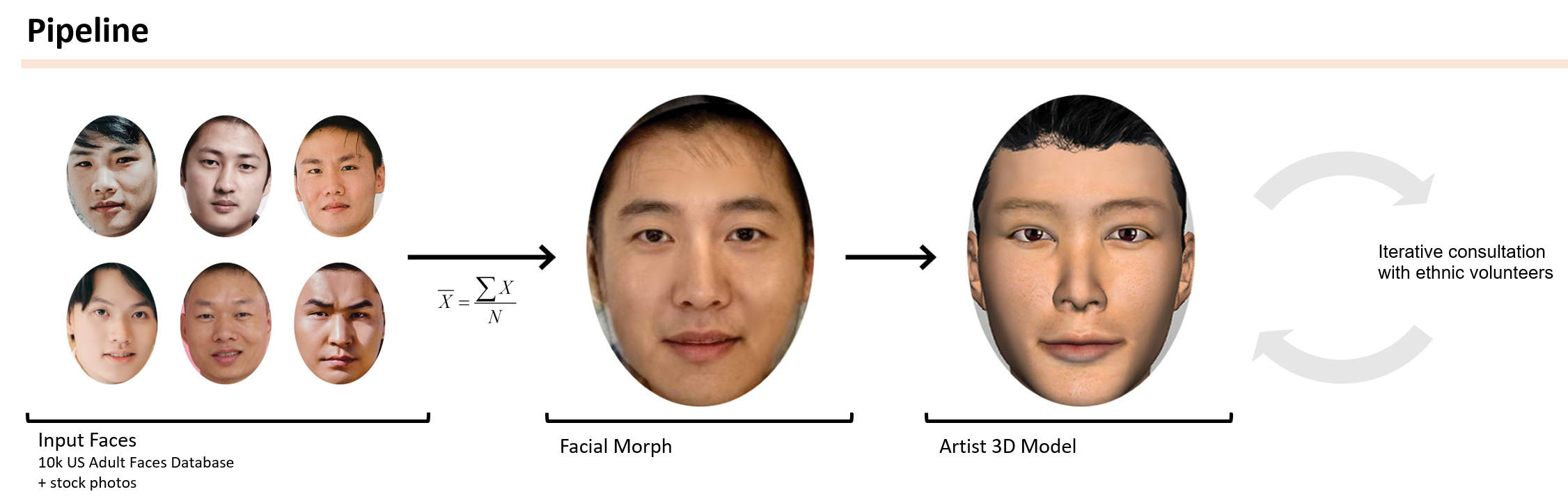}
    \caption{An example of the creation of a 3D avatar using our methodology. 1) We select 4-7 faces from a database \cite{Bainbridge2013} or stock photos. 2) We calculate the average face using WebMorph \cite{DeBruine2018}. 3) A 3D artist recreates the average face using modeling software. 4) The models are improved iteratively through recurrent consultation with representative volunteers.}
    \label{fig:averages}
\end{figure*}
This section outlines our iterative process for developing the VALID library, which includes 42 base avatars. We began by using data-driven averaged facial features to create our initial models. We then conducted interviews with representative volunteers to iteratively refine and modify the avatars based on their feedback.

\subsection{Initial Modeling}
To ensure a broad diversity of people were represented in our library, we initially created 42 base avatars (7 target races $\times$ 2 genders $\times$ 3 individuals ) modeled after the seven racial groups recommended by the 2015 National Content Test Race and Ethnicity Analysis Report \cite{Matthews2017}: AIAN, Asian, Black, Hispanic, MENA, NHPI, and White. We created 3 male and 3 female individuals for each race, resulting in a total of 6 individuals per race.

Preliminary models were based on averaged facial features of multiple photos selected from the 10k US Adult Faces Database \cite{Bainbridge2013} and stock photos from Google for races missing from the database (e.g., AIAN, MENA, and NHPI). These photos were used as input to a face-averaging algorithm \cite{DeBruine2018}, which extracted average facial features for each race and gender pair. Using these averages as a reference, a 3D artist recreated the average faces for each race and gender pair using Autodesk Character Generator (due to its generous licensing and right to freely edit and distribute generated models\footnote{https://knowledge.autodesk.com/support/character-generator/learn-explore/caas/sfdcarticles/sfdcarticles/Character-Generator-Legal-Restrictions-and-Allowances-when-using-Character-Generator.html}) and Blender to make modifications not supported by Autodesk Character Generator (see Figure \ref{fig:averages}). 

\subsection{Iterative Improvements through Representative Interviews}
After the preliminary avatars were created based on the facial averages, we worked closely with 2 to 4 volunteers of each represented race (see Table \ref{table:representatives}) to adjust the avatars through a series of Zoom meetings. This process ensured that all avatars were respectful and reduced the likelihood of harmful or stereotypical representations. Volunteers self identified their race and were recruited from university cultural clubs (e.g., Asian Student Association, Latinx Student Association), community organizations (e.g., Pacific Islanders Center), and email lists. We iteratively asked these volunteers for feedback on all avatars representing their race, showing them the model from three perspectives (see Figure \ref{fig:screenshots}). Volunteers were specifically asked to identify accurate features and suggest changes to be made. Once the changes were completed based on the feedback, we presented the updated avatars to the volunteers. This process was repeated until they approved the appearance of the avatars. For example, volunteers requested changes to facial features, such as:

\begin{itemize}
    \item \textit{``Many Native women I know have a softer face and jawline [than this avatar].''} --AIAN Volunteer 3
    \item \textit{``The nose bridge is too high and looks too European. Asians mostly have low nose bridges.''} --Asian Volunteer 2
    \item \textit{``Middle Eastern women usually have wider, almond-shaped eyes.''} --MENA Volunteer 2
    \item \textit{``The nose [for this avatar] should be a little bit thicker, less pointy, and more round.''} --NHPI Volunteer 1
\end{itemize}

Additionally, we modified hairstyles according to feedback: 

\begin{itemize}
    \item \textit{``[These hairstyles] look straighter and more Eurocentric. So I would choose [these facial features] and then do a natural [hair] texture.''} --Black Volunteer 1
    \item \textit{``Usually the men have curly hair or their hair is cut short on the sides with the top showing.''} --NHPI Volunteer 1
\end{itemize}

Once the avatars were approved by their corresponding volunteer representatives, we conducted an online study to validate the race and gender of each avatar based on user perceptions.

\begin{table}[h] \small
\caption{Breakdown of our volunteer representatives by race, gender (male, female, or non-binary), and country.}
\begin{tabular}{@{}lll@{}}
\toprule
\textbf{Race} & \textbf{Gender} & \textbf{Country} \\ \midrule
\textbf{AIAN} & 2M, 1F & USA (Native American) (3) \\
\rowcolor[HTML]{EFEFEF} 
\textbf{Asian} & 2M, 2F & \begin{tabular}[c]{@{}l@{}}China (1), USA (Chinese-American) (1),\\ USA (Vietnamese-American) (2)\end{tabular} \\
\textbf{Black} & 1M, 2F & USA (African-American) (3) \\
\rowcolor[HTML]{EFEFEF} 
\textbf{Hispanic} & 2M, 1F & Mexico (1), USA (Mexican-American) (2) \\
\textbf{MENA} & 1M, 2F & Iran (2), Saudi Arabia (1) \\
\rowcolor[HTML]{EFEFEF} 
\textbf{NHPI} & 1M, 2F & Samoa (2), USA (Native Haiwaiian) (1) \\
\textbf{White} & 2M, 1F & USA (3) \\ \bottomrule
\end{tabular}
\label{table:representatives}
\end{table}

\section{Avatar Validation Study}
We conducted an online, worldwide user study to determine whether the target race and gender of each avatar is recognizable and, therefore, validated. Participants were recruited from the online Prolific marketplace\footnote{https://www.prolific.co}, which is similar to Amazon Mechanical Turk. Prior research shows that Prolific has a pool of more diverse and honest participants \cite{Peer2017} and has more transparency than Mechanical Turk \cite{Palan2018}. Since diversity was a core theme of our research, we chose Prolific to ensure that our participants would be diverse.

\subsection{Procedure}
The following procedure was reviewed and approved by our university Institutional Review Board (IRB). The study consisted of one online Qualtrics survey that lasted an average of 14 minutes. Each participant first completed a background survey that captured their self-identified demographics, including race, gender, and education. Afterwards, they were asked to familiarize themselves with the racial terms as defined by the U.S. Census Bureau research \cite{Matthews2017}. Participants were then asked to categorize the 42 avatars by their perceived race and gender. Participants were shown only one avatar at a time and the order was randomized. 

\begin{figure}[]
    \centering
    \includegraphics[width=3in]{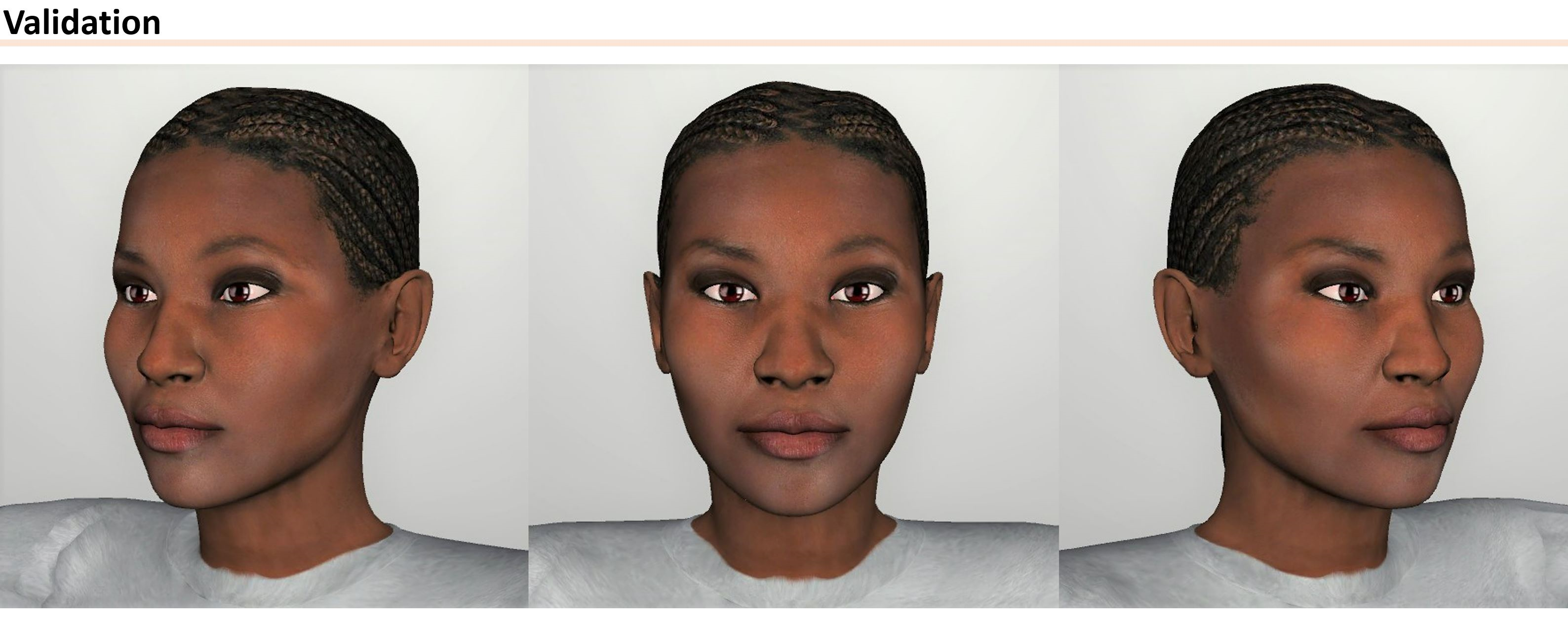}
    \caption{An example of how each avatar was presented to participants during our validation study.}
    \label{fig:screenshots}
\end{figure}

For each of the avatars, participants were shown three perspectives: a $45^{\circ}$ left headshot, a direct or $0^{\circ}$ headshot, and a $45^{\circ}$ right headshot (see Figure \ref{fig:screenshots}). Avatars were shown from the shoulders up and were dressed in a plain gray shirt. The images were rendered in Unity using the standard diffuse shader and renderer. The avatars were illuminated by a soft white (\#FFFEF5) directional light with an intensity of 1.0, and light gray (\#7F7F7F) was used for the background. Participants were asked to select all races that each avatar could represent: ``American Indian or Alaskan Native'', ``Asian'', ``Black or African American'', ``Hispanic, Latino, or Spanish'', ``Middle Eastern or North African'', ``Native Hawaiian or Pacific Islander'', ``White'', or ``Other''. ``Other'' included an optional textbox if a participant wanted to be specific. We allowed participants to select multiple categories according to the U.S. Census Bureau's recommendations for surveying race \cite{Matthews2017}. For gender, participants were able to select ``Male'', ``Female'', or ``Non-binary''. Participants were paid \$5.00 via Prolific for completing the study. 

\subsection{Participants}
A total of 132 participants (65 male, 63 female, 4 non-binary) from 33 different countries were recruited to take part in the study. We aimed to ensure a diverse representation of perspectives by balancing participants by race and gender. Table \ref{table:participants} provides a breakdown of our participants by race, gender, and country. Despite multiple recruitment attempts, including targeted solicitations via Prolific, we had difficulty recruiting NHPI participants. It is important to note that we excluded volunteers who had previously assisted with modeling the avatars from participating in the validation study to avoid potentially overfitting their own biases.

\begin{table}[h!] \small 
\caption{Breakdown of our validation study's participants by race, gender (male, female, or non-binary), and country.}
\begin{tabular}{@{}lll@{}}
\toprule
\textbf{Race}  & \textbf{Gender} & \textbf{Country} \\ \midrule
\textbf{AIAN} & 9M, 10F, 1NB & \begin{tabular}[c]{@{}l@{}}U.S. (14), Canada (3), Chile (1), Mexico (1), \\Cambodia (1) \end{tabular}\\
\rowcolor[HTML]{EFEFEF} 
\textbf{Asian} & 10M, 10F & \begin{tabular}[c]{@{}l@{}}U.K. (5), Canada (3), South Africa (3), \\India (2), Indonesia (2), China (1), France (1), \\Germany (1), Italy (1), Malaysia (1)\end{tabular} \\
\textbf{Black} & 10M, 9F, 1NB & \begin{tabular}[c]{@{}l@{}}South Africa (16), Nigeria (2), \\Swaziland (1), U.K. (1) \end{tabular}\\
\rowcolor[HTML]{EFEFEF} 
\textbf{Hispanic} & 10M, 9F, 1NB & Mexico (15), Chile (3), Portugal (2) \\
\textbf{MENA} & 10M, 10F & \begin{tabular}[c]{@{}l@{}}Israel (9), Lebanon (3), Jordan (2), \\Bahrain (1), Egypt (1), Iran (1), Iraq (1), \\Saudi Arabia (1), Syria (1)\end{tabular} \\
\rowcolor[HTML]{EFEFEF} 
\textbf{NHPI} &  7M, 5F & \begin{tabular}[c]{@{}l@{}}New Zealand (Maori) (8), Samoa (2),\\ Fiji (1), U.S. (1)\end{tabular}\\
\textbf{White} & 9M, 10F, 1NB & \begin{tabular}[c]{@{}l@{}}Portugal (5), Italy (5), Poland (3), \\Mexico (2), Belgium (1), Greece (1), \\Ireland (1), U.K. (1), U.S. (1)\end{tabular} \\ \bottomrule
\end{tabular}
\label{table:participants}
\end{table}

\subsection{Data Analysis and Labeling Approach}
To validate the racial identification of our virtual avatars, we used Cochran's Q test \cite{Sheskin2011}, which allowed us to analyze any significant differences among the selected race categories. This approach was necessary since our survey format allowed participants to select more than one race category for each avatar, following the U.S. Census Bureau's research recommendations \cite{Matthews2017}. Since the Chi-squared goodness of fit test requires mutually exclusive categories, we were unable to use it in our analysis. Furthermore, since our data was dichotomous, a repeated-measures analysis of variance (ANOVA) was not appropriate. Therefore, Cochran's Q test was the most appropriate statistical analysis method for our survey data.

We used a rigorous statistical approach to assign race and gender labels to each avatar. First, we conducted the Cochran's Q test across all participants ($n=132$) at a 95\% confidence level to identify significant differences in the participants' responses. If the test indicated significant differences, we performed pairwise comparisons between each race using Dunn's test to determine which races were significantly different. 

For each avatar, we assigned a race label if the race was selected by the majority of participants (i.e., over 50\% of participants selected it) and if the race was selected significantly more than other race choices and not significantly less than any other race. This approach resulted in a single race label for most avatars, but some avatars were assigned multiple race labels due to multiple races being selected significantly more than all other races. If no race was selected significantly more than the majority, then we categorized the avatar as ``Ambiguous''. We followed a similar procedure for assigning gender labels.

To account for the possibility that the race of the participant might influence their perception of virtual race, we also assigned labels based on same-race participants. This involved using the same procedure for assigning labels as described above, except based only on the selections of participants who identified as the same race as the avatar. This also allows future researchers to have the flexibility to use the labels from all study participants for studies focused on individuals from diverse racial backgrounds or to use the labels from participants of the same race for studies targeting specific racial groups.

\section{Results}

\begin{table*}[h] \small
\caption{Assigned labels for all 42 base avatars. ``All'' indicates that the label was identified by all 132 participants, while ``Same-Race'' only includes the data of participants who identify as the race that the avatar was modeled for. Agreement labels were calculated as the percentage of participants who perceived an avatar to represent a race or gender.}
\begin{tabular}{@{}lllllll@{}}
\toprule
\textbf{Avatar} & \textbf{Race (All)} & \textbf{Agreement} & \textbf{Race (Same-Race)} & \textbf{Agreement} & \textbf{Gender (All)} & \textbf{Agreement} \\ \midrule
\textbf{AIAN\_M\_1} & AIAN & 0.67 & AIAN & 0.90 & Male & 0.89 \\
\textbf{AIAN\_M\_2} & AIAN & 0.65 & AIAN & 0.75 & Male & 0.89 \\
\textbf{AIAN\_F\_1} & AIAN & 0.55 & AIAN & 0.60 & Female & 0.64 \\
\textbf{AIAN\_F\_2} & AIAN & 0.55 & AIAN & 0.75 & Female & 0.75 \\
\textbf{Asian\_M\_1} & Asian & 0.86 & Asian & 0.90 & Male & 0.95 \\
\textbf{Asian\_M\_2} & Asian & 0.98 & Asian & 1.00 & Male & 0.91 \\
\textbf{Asian\_M\_3} & Asian & 0.98 & Asian & 1.00 & Male & 0.91 \\
\textbf{Asian\_F\_1} & Asian & 0.98 & Asian & 1.00 & Female & 0.98 \\
\textbf{Asian\_F\_2} & Asian & 0.93 & Asian & 0.90 & Female & 0.96 \\
\textbf{Asian\_F\_3} & Asian & 0.95 & Asian & 1.00 & Female & 0.97 \\
\textbf{Black\_M\_1} & Black & 0.98 & Black & 1.00 & Male & 0.98 \\
\textbf{Black\_M\_2} & Black & 0.97 & Black & 1.00 & Male & 0.98 \\
\textbf{Black\_M\_3} & Black & 0.98 & Black & 1.00 & Male & 0.73 \\
\textbf{Black\_F\_1} & Black & 0.98 & Black & 1.00 & Female & 0.98 \\
\textbf{Black\_F\_2} & Black & 0.98 & Black & 1.00 & Female & 0.94 \\
\textbf{Black\_F\_3} & Black & 0.95 & Black & 1.00 & Female & 0.97 \\
\textbf{Hispanic\_M\_1} & Hispanic/White & 0.65/0.55 & Hispanic & 0.85 & Male & 0.98 \\
\textbf{Hispanic\_M\_2} & White & 0.64 & Hispanic & 0.80 & Male & 0.98 \\
\textbf{Hispanic\_M\_3} & Hispanic/White & 0.55/0.51 & Hispanic & 0.75 & Male & 0.97 \\
\textbf{Hispanic\_F\_1} & Hispanic & 0.64 & Hispanic & 0.80 & Female & 0.95 \\
\textbf{Hispanic\_F\_2} & White & 0.70 & Hispanic/White & 0.55/0.60 & Female & 0.95 \\
\textbf{Hispanic\_F\_3} & Hispanic/White & 0.59/0.52 & Hispanic & 0.75 & Female & 0.93 \\
\textbf{MENA\_M\_1} & White & 0.56 & MENA & 0.70 & Male & 0.99 \\
\textbf{MENA\_M\_2} & White & 0.64 & MENA/White & 0.65/0.60 & Male & 1.00 \\
\textbf{MENA\_M\_3} & MENA/White & 0.55/0.65 & MENA/White & 0.75/0.55 & Male & 0.98 \\
\textbf{MENA\_F\_1} & White & 0.58 & MENA/White & 0.70/0.60 & Female & 0.98 \\
\textbf{MENA\_F\_2} & White & 0.60 & MENA & 0.60 & Female & 0.98 \\
\textbf{NHPI\_M\_1} & NHPI & 0.52 & NHPI & 0.58 & Male & 0.98 \\
\textbf{NHPI\_M\_2} & Hispanic & 0.65 & NHPI & 0.67 & Male & 1.00 \\
\textbf{White\_M\_1} & White & 0.96 & White & 0.95 & Male & 0.99 \\
\textbf{White\_M\_2} & White & 0.98 & White & 0.95 & Male & 1.00 \\
\textbf{White\_M\_3} & White & 0.93 & White & 0.90 & Male & 0.99 \\
\textbf{White\_F\_1} & White & 0.94 & White & 0.95 & Female & 0.97 \\
\textbf{White\_F\_2} & White & 0.96 & White & 0.95 & Female & 0.98 \\
\textbf{White\_F\_3} & White & 0.94 & White & 0.95 & Female & 0.99 \\
\textbf{X\_AIAN\_M\_1} & Hispanic & 0.64 & Hispanic & 0.75 & Male & 0.99 \\
\textbf{X\_AIAN\_F\_1} & Hispanic & 0.54 & Hispanic & 0.45 & Female & 0.86 \\
\textbf{X\_MENA\_F\_1} & Ambiguous & N/A & Ambiguous & N/A & Female & 0.98 \\
\textbf{X\_NHPI\_M\_1} & Hispanic & 0.55 & Asian & 0.58 & Male & 0.98 \\
\textbf{X\_NHPI\_F\_1} & Hispanic & 0.55 & Ambiguous & N/A & Female & 0.92 \\
\textbf{X\_NHPI\_F\_2} & Hispanic & 0.58 & Ambiguous & N/A & Female & 0.95 \\ 
\textbf{X\_NHPI\_F\_3} & NHPI & 0.52 & Ambiguous & N/A & Female & 0.92 \\
\midrule
\multicolumn{7}{l}{\begin{tabular}[c]{@{}l@{}}
\end{tabular}}
\end{tabular}
\label{table:labels}
\end{table*}

\subsection{Validated Avatar Labels}
Table \ref{table:labels} summarizes our results and labels for all 42 base avatars across all participants and for same-race participants. 

\subsubsection{Race and Gender Labels}
Asian, Black, and White avatars were correctly identified as their intended race across all participants, while most of the remaining avatars were accurately identified by same-race participants (see Table \ref{table:labels} for all and same-race agreement rates). Therefore, we observed some differences in identification rates based on the race of the participants, highlighting the potential impact of own-race bias on the perception of virtual avatars. Notably, there were no significant differences in gender identification rates based on participant race, indicating that all avatars were correctly perceived as their intended gender by all participants, regardless of their racial background. 

\subsubsection{Naming Convention}
If an avatar was identified as its intended race by corresponding same-race participants, we named it after that race. For instance, the avatar \verb|Hispanic_M_2| was labeled as White by all participants. However, our Hispanic participants perceived it as solely Hispanic. Hence, we left the original name. However, if an avatar was labeled as ``Ambiguous'' or as a different race by same-race participants, we added an \verb|X| at the beginning of its name to indicate that it was not validated. Avatars were also labeled by their identified gender (``M" or ``F").

\begin{figure*}[h!]
    \centering
    \includegraphics[width=6.8in]{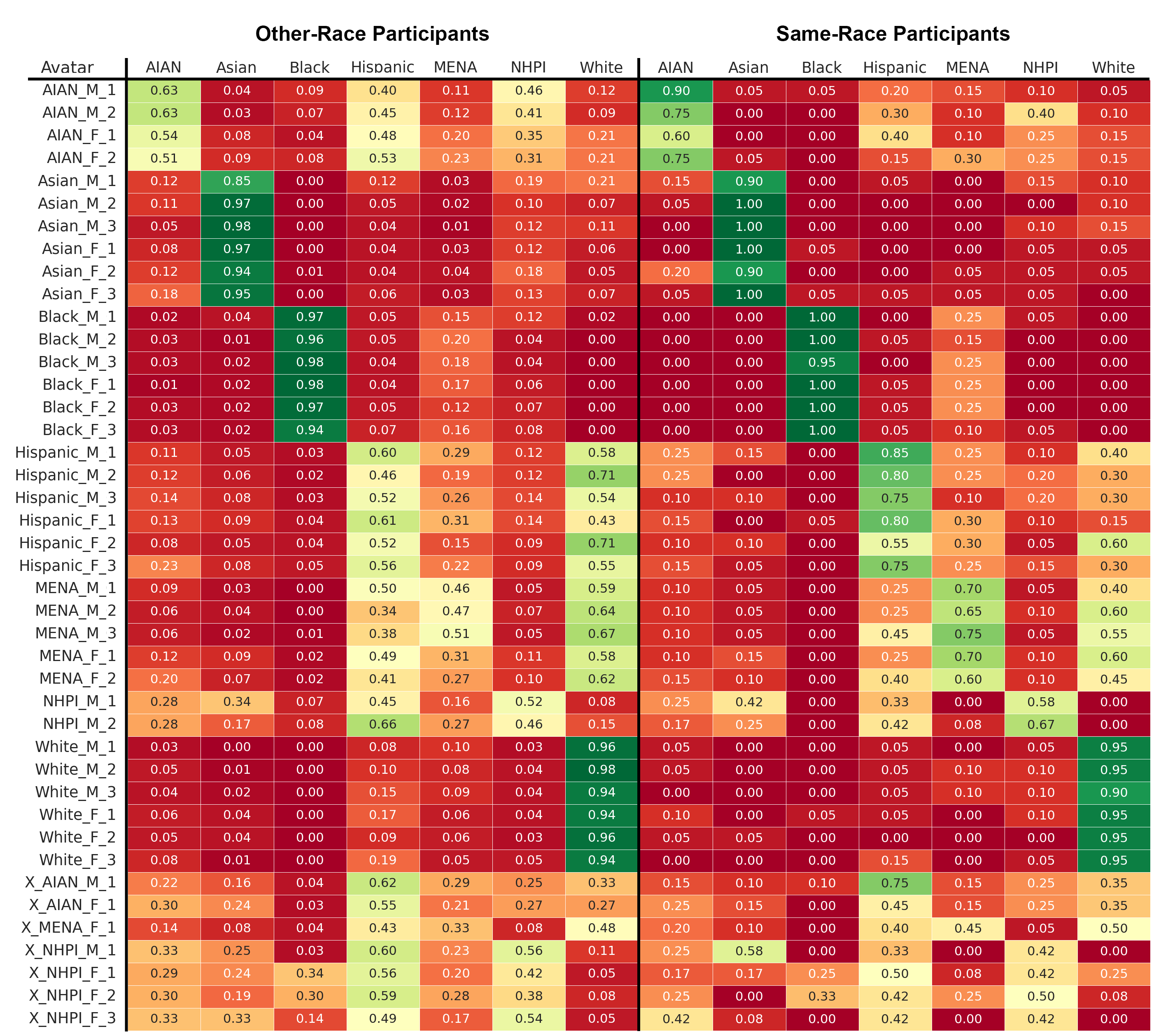}
    \caption{Confusion matrix heatmap of agreement rates for the 42 base avatars by separated by other-race participants and same-race participants (i.e., participants of a different or same race as the avatar). Agreement rates were calculated as the percentage of participants who perceived an avatar to represent a race or gender.}
    \label{fig:agreement_all}
\end{figure*}

\subsection{Other-Race vs. Same-Race Perception}
To further examine how participant race affected perception of virtual avatar race, we additionally analyzed the data by separating same-race and other-race agreement rates. In effect, we separated the selections of the participants who were the same race as the avatar modeled and those who were not. 

\subsubsection{Difference in Agreement Rates}
 Figure \ref{fig:agreement_all} displays the difference in agreement rates between same-race and other-race. Figure \ref{fig:agreement_all} shows that several avatars were strongly identified by both other-race and same-race participants. In particular, all Asian, Black, and White avatars were perceived as their intended race with high agreement rates by both same-race and other-race participants (over 90\% agreement for all but one). However, some avatars were only identified by participants of the same race as the avatar. For example, our analysis of the agreement rates for different racial groups revealed interesting trends. For instance, non-Hispanic participants had an average agreement rate of 54.5\% for Hispanic avatars, while Hispanic participants had a much higher average agreement of 75.0\%. Similar patterns were observed for AIAN (57.8\% other-race, 75.0\% same-race) and MENA (40.4\% other-race, 68.0\% same-race) avatars.

\begin{figure*}[ht!]
\centering
\subfigure[Other-race clustering]{
\includegraphics[width=.45\textwidth]{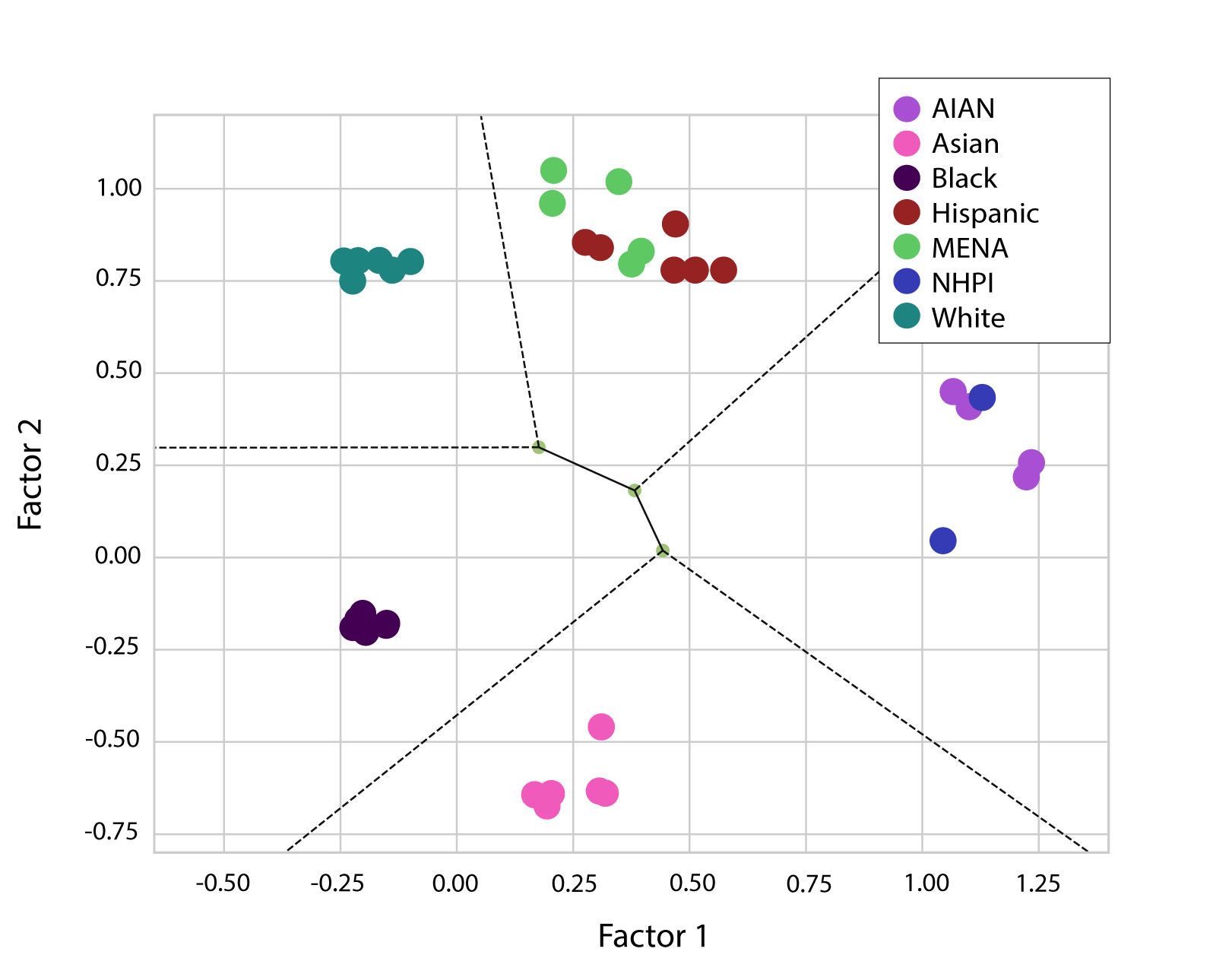}
}
\subfigure[Same-race clustering]{
\includegraphics[width=.45\textwidth]{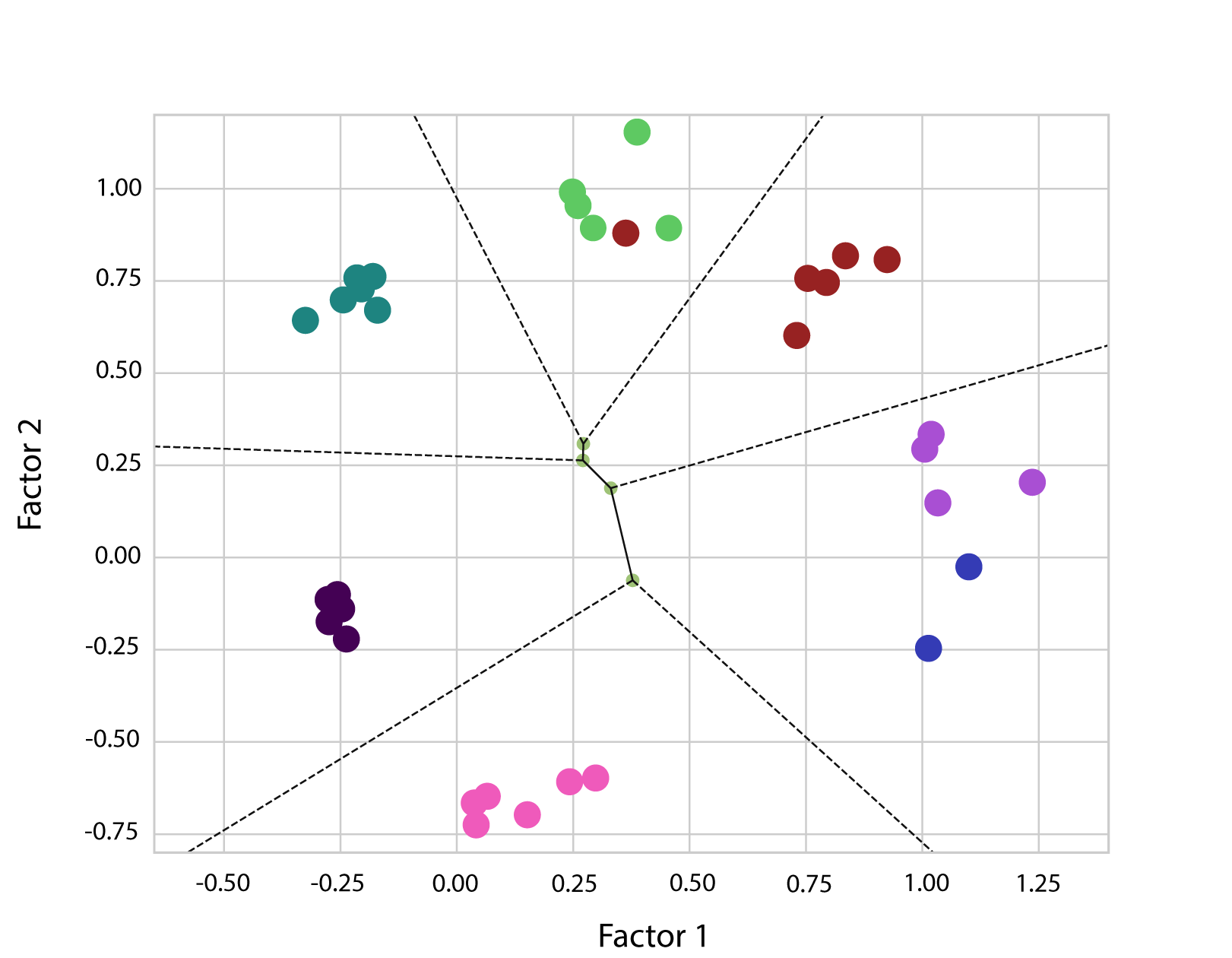}
}

\caption{Clustered scatterplots of each avatar's relation to one another based on Principle Component Analysis and K-means clustering for other-race and same-race participant identifications. The Voronoi analysis shows the borders of the clusters where each category was assigned. Each avatar is color coded by its validated label.}
\label{fig:scatterplot}
\end{figure*}

\subsubsection{Perceived Race Clusters}
To gain deeper insights into how participants perceived the avatars' races, we employed Principle Component Analysis (PCA) to reduce the agreement rates of each of the 42 base avatars down to two dimensions. Next, we performed K-means clustering \cite{Kmeans} on the resulting two-dimensional data to group the avatars based on their perceived race. We optimized the number of clusters using the elbow method and distortion scores \cite{bengfort_yellowbrick_2018}. We applied this technique to both other-race and same-race agreement rates to determine whether there were any differences in the clustering based on participant race. By visualizing the clusters, we aimed to better understand the differences in how participants perceived the avatars' races.

Figure \ref{fig:scatterplot} shows that Asian, Black, and White avatars were perceived consistently by all participants, with clearly defined clusters. However, there was more confusion in perceiving AIAN, Hispanic, MENA, and NHPI avatars, which clustered closer together. Same-race participants had less overlap and more-accurately perceived these avatars, with more separation between them. For example, the Hispanic and MENA avatars were in separate clusters for same-race participants, except for one avatar (\verb|Hispanic_F_2|). On the other hand, the Hispanic and MENA avatars were entirely clustered together for other-race participants. 

\section{Discussion}
In this section, we discuss the validation of our avatars. Specifically, we examine the extent to which each avatar was correctly identified as its intended race and the variability in identification across different participant groups. Additionally, we discuss the implications of our results for virtual avatar research, highlighting the importance of considering the potential impact of own-race bias on avatar race perception. Finally, we describe the potential future impact of our avatar library in the community, including how it can be used to promote diversity and inclusion. 

\begin{figure*}[h]
    \centering
    \includegraphics[width=6.9in]{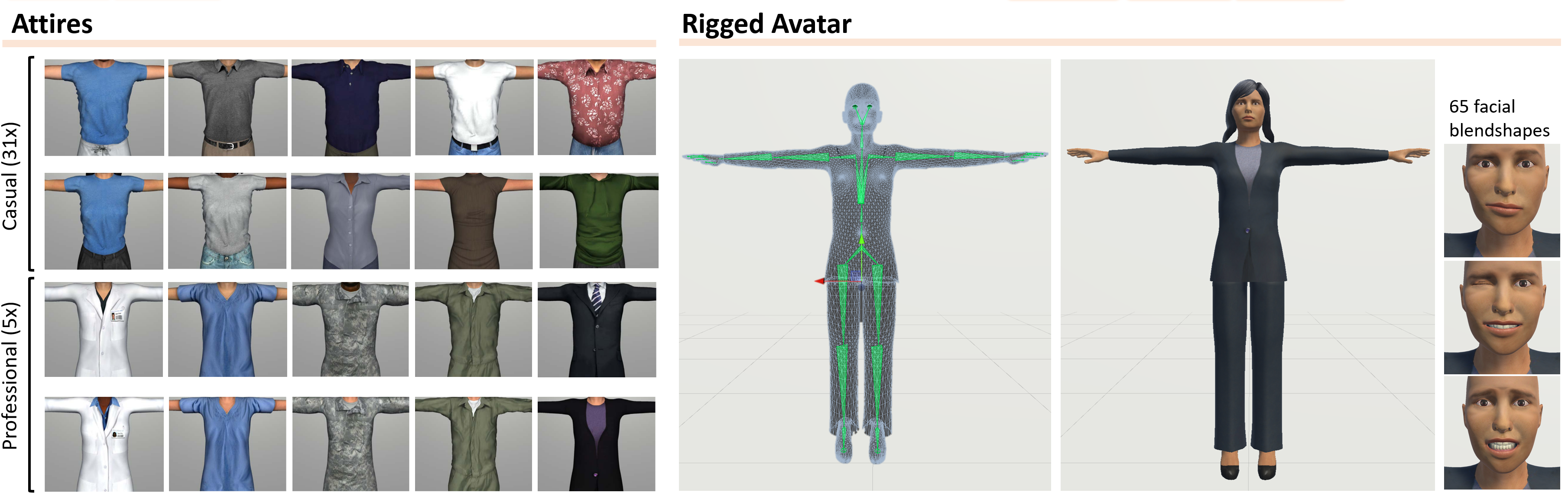}
    \caption{Images of the skeleton and facial blend shapes included with our avatars.}
    \label{fig:skeleton}
\end{figure*}

\subsection{Race Identification}
\subsubsection{Universally Identified Avatars}
We found that our Asian, Black, and White avatars were recognized by all participants with high agreement rates. This suggests that these avatars can be a valuable tool for researchers seeking to create virtual humans that can be easily identified by individuals from different racial backgrounds.

Our results may be due to perceptual expertise or familiarity with other-race faces, as proposed by Civile et al. \cite{Civile2022}. We hypothesize that this familiarity could be explained by the prevalence of these racial groups in global media and pop culture. For example, White cast members were the most represented in popular Hollywood movies over the last decade, followed by Black cast members \cite{Malik2022}. Since Hollywood movies have a dominant share in the global film industry \cite{Maisuwong2012}, people may be more familiar with characters that are prevalent in these films. Additionally, East Asian media culture has become widely popular worldwide over the past few decades \cite{jin2021, Iwabuchi2010}. Phenomena like "The Hallyu Wave" and "Cool Japan" \cite{Lux2021} have enabled East Asian films, dramas, and pop music to gain a global following. As people may often encounter these racial groups in media, this familiarity may have facilitated their recognition of these avatars.

\subsubsection{Same-Race Identified Avatars} \label{WithinEthnicity}
As expected, some avatars were only identified by participants of the same race as the avatar, consistent with the own-race bias effect. For example, as seen in Table \ref{table:labels}, the Hispanic avatars received mixed ratings of White and Hispanic across all participants, but most were perceived as solely Hispanic by Hispanic-only participants. Similarly, only one MENA avatar was perceived as MENA by all participants, while five were perceived as MENA by MENA-only participants. These results suggest that participants' own-race bias, a well-known phenomenon in psychology, may also affect their perception of virtual avatars. The findings point to the importance of considering participants' race when using virtual avatars in research or applications that require accurate representation of different racial groups.

\subsubsection{Ambiguous Avatars}
Several avatars in our library were perceived ambiguously by all participants and only same-race participants, and therefore labeled as such (see Table \ref{table:labels} for details). Identifying the reason for these avatars' lack of clear identification is not straightforward, and multiple factors could be at play. For instance, the two ambiguous AIAN avatars were the only ones with short hairstyles, which may have impacted their identification as AIAN. Long hair carries cultural and spiritual significance in many AIAN tribes \cite{Thanikachalam2019}, and some participants may have perceived the avatars as non-AIAN as a result, even among AIAN participants.

The validation of our NHPI avatars was limited, possibly due to the low number of NHPI participants ($n=12$) in our study, despite our targeted recruitment efforts. As a consequence, most of the NHPI avatars were not validated by NHPI participants, including the lack of validation for any female NHPI avatars. Another potential reason for this lack of validation is that the majority of our NHPI participants identified themselves as New Zealand Maori, whereas our avatars were developed with the help of Samoan and Native Hawaiian volunteer representatives. Therefore, it is possible that our NHPI avatars are representative of some NHPI cultures, but not New Zealand Maori. In future studies, expanding recruitment efforts for both interview volunteers and study participants will be crucial, despite the challenges involved in doing so. For example, future studies may need to compensate NHPI participants more than participants of other races.

\subsection{Implications for Virtual Avatars}
Our study provides valuable insights for virtual avatar applications and research. Our findings indicate that human behavior in race categorization can apply to virtual avatars, which has notable implications for interactions in virtual experiences. Kawakami et al. \cite{Kawakami2017} suggest that in-group and out-group categorization can lead to stereotyping, social judgments, and group-based evaluations. Therefore, designers and developers should be aware of this and take necessary steps to mitigate unintended consequences in virtual experiences. For example, regulating codes of conduct \cite{Jones2020} can help to improve interracial interactions in VR.

Interestingly, our study also replicated a nuanced finding from more recent psychology research on the perception of ambiguous avatars \cite{Nicolas2019}. As seen in Table \ref{table:labels}, most of the misidentified avatars were identified as Hispanic by all participants. Similarly, Nicolas et al. \cite{Nicolas2019} recently found that participants classify racially ambiguous photos as Hispanic or MENA, regardless of their parent ethnicities. We believe that this effect extended to our virtual avatars.

\subsection{An Open Library of Validated Avatars}
As a contribution to the research community, we are providing open access to our virtual avatar library, which includes all 210 fully rigged avatars, along with validated labels for each avatar's race and gender. Our library features avatars of seven different races, providing a diverse selection for researchers to use in their studies. The validated labels can facilitate research on the impact of avatar race, and researchers can choose to use the labels for studies aimed at individuals from different racial backgrounds or same-race labels for specific study populations.

The \textit{Virtual Avatar Library for Inclusion and Diversity} (\textit{VALID}) provides researchers and developers with a diverse set of fully rigged avatars suitable for various scenarios such as casual, business, medical, military, and utility. Each avatar comes with 65 facial blend shapes, enabling dynamic facial expressions (see Figure \ref{fig:skeleton}). The library is readily available for download and can be used in popular game engines like Unity or Unreal. Although this is the first iteration of the library, we plan to update it by adding more professions and outfits soon. In addition, the library can be used for a wide range of research purposes, including social psychology simulations and educational applications.

\subsection{Limitations and Future Work}
We recognize that our VALID avatar library is only a small step towards achieving greater diversity and inclusion in avatar resources. We acknowledge that the representation of each demographic is limited and plan to expand the diversity within each group by creating new avatars. For example, our Asian avatars are modeled after East Asians, but we plan to expand VALID to include South Asian and Southeast Asian avatars as well. Our Hispanic representatives have pointed out the need for more diverse Hispanic avatars, including varying skin tones to represent different South American populations, such as Mexican and Cuban. Additionally, our NHPI representatives have suggested the inclusion of tattoos, which hold cultural significance for some NHPI communities, could improve the identifiability of our NHPI avatars, in addition to improving our NHPI recruitment methods. Any future updates to the library will undergo the same rigorous creation, iteration, and validation process as the current avatars.

While our first iteration of the library focused on diversity in terms of race, we realize that the avatars mostly represent young and fit adults, which does not reflect all types of people. In the future, we plan to update the library with a diversity of body types that include different body mass index (BMI) representations and ages. Including avatars with different BMI representations is not only more inclusive, but can also be useful for studies targeting physical activity, food regulation, and therapy \cite{Scarpina2019}. Likewise, we plan to include shaders and bump maps \cite{Hughes2011} that can age any given avatar by creating realistic wrinkles and skin folds, further improving the diversity and inclusivity of VALID

Another limitation of the current work is that our library includes only male and female representations. In future updates, we plan to include non-binary and androgynous avatars. Currently, there are not many androgynous models that are freely available. However, they can be an area of important study. For example, previous studies found that androgynous avatars reduce gender bias and stereotypical assumptions in virtual agents \cite{Nag2020} and improve student attitudes \cite{Gulz2010}. Thus, we plan to include these avatars in a future update by following Nag et al.'s \cite{Nag2020} guidelines for creating androgynous virtual humans. 

Our study, while diverse in terms of race and country, is not representative of everyone. We recruited participants through the online platform Prolific, which is known for its increased diversity compared to other crowdsourcing platforms such as Mechanical Turk. However, due to the online nature of the platform, we primarily recruited younger adults. It is possible that perceptions of our avatars may differ among other age groups, such as children or older adults. Therefore, it is important to broaden recruitment efforts by exploring alternative platforms and recruitment strategies that may be more effective in reaching a wider range of participants. Future studies could also consider conducting in-person studies or focus groups to gather additional insights into avatar perception.

\section{Conclusion}
We have introduced a new virtual avatar library comprised of 210 fully rigged avatars with diverse professions and outfits, available for free. Our library aims to promote diversity and inclusion by creating equitable representation of seven races across various professions. We designed 42 base avatars using data-driven facial averages and collaborated with volunteer representatives of each ethnicity. A large validation study involving participants from around the world was conducted to obtain validated labels and metadata for the perceived race and gender of each avatar. Additionally, we offer a comprehensive process for creating, iterating, and validating diverse avatars to aid other researchers in creating similarly validated avatars.

Our validation study revealed that the majority of avatars were accurately perceived as the race they were modeled for. However, we observed that some avatars, such as the Hispanic and MENA avatars, were only validated as such by participants who identified as Hispanic or MENA, respectively. This finding suggests that the perception of virtual avatars may be influenced by own-race bias or the other-race effect, as described in the psychology literature. Moving forward, we plan to expand the library to include additional races, professions, body types, age ranges, and gender representations to further improve diversity and inclusion.
\bibliographystyle{ACM-Reference-Format}
\bibliography{sample-base}

\newpage
\setcounter{figure}{0}
\renewcommand{\thefigure}{A\arabic{figure}}

\end{document}